\documentclass[a4paper,12pt]{article}
\usepackage{amssymb,color}
\usepackage{graphicx}
\usepackage{apacite}

\newtheorem{algo}{Algorithm}

\newtheorem{proposition}{Proposition}

\def\Z {\textbf{Z}}

\def\X {\textbf{X}}

\def\p {^{\prime}}
\def\t {\theta}

\def\b {\beta}

\def\s {\sigma^2}
\def\d {\delta}

\begin{document}

\title{Random Indicator Imputation for Missing Not At Random Data}
\author{S. Jolani$^\dag$\thanks{Corresponding author. Email: s.jolani@maastrichtuniversity.nl}, S. van Buuren$^{\ddag}$\\
{\footnotesize\it\dag Department of Methodology and Statistics, Maastricht University, Maastricht, The Netherlands}\\
{\footnotesize\it\ddag Department of Statistics, TNO, Leiden, The Netherlands}
}
\date{}
\maketitle

\begin{abstract}
Imputation methods for dealing with incomplete data typically assume that the missingness mechanism is at random (MAR). These methods can also be applied to missing not at random (MNAR) situations, where the user specifies some adjustment parameters that describe the degree of departure from MAR. The effect of different pre-chosen values is then studied on the inferences. This paper proposes a novel imputation method, the Random Indicator (RI) method, which, in contrast to the current methodology, estimates these adjustment parameters from the data. For an incomplete variable $X$, the RI method assumes that the observed part of $X$ is normal and the probability for $X$ to be missing follows a logistic function. The idea is to estimate the adjustment parameters by generating a pseudo response indicator from this logistic function. Our method iteratively draws imputations for $X$ and the realization of the response indicator $R$, to which we refer as $\dot{R}$, for $X$. By cross-classifying $X$ by $R$ and $\dot{R}$, we obtain various properties on the distribution of the missing data. These properties form the basis for estimating the degree of departure from MAR. Our numerical simulations show that the RI method performs very well across a variety of situations. We show how the method can be used in a real life data set. The RI method is automatic and opens up new ways to tackle the problem of MNAR data.


\bigskip
\textit{\textbf{keywords:}} Multiple imputation; Nonresponse model; Propensity score; Sensitivity analysis;
\end{abstract}

\section{Introduction}
Missing values are a common problem in the statistical analysis of data in all disciplines including medical or social studies. These may occur because the intended measurements are not taken, lost, or unavailable. Two major consequences of missing data are loss of precision and possibility of introducing bias. Statisticians can do little about the former while they can aim to reduce the later by using an appropriate analysis.

Any missing data analysis makes assumptions about the mechanisms that produce missing data. \citeA{rubin76} introduced taxonomy of the missingness mechanisms: Missing Completely At Random (MCAR), Missing At Random (MAR), and Missing Not At Random (MNAR). Methods for handling missing data often assume MCAR or MAR. The degree to which these assumptions are violated can have serious impact on the validity of the analysis. A less restricted assumption is MNAR where the mechanism governing missingness depends on unobserved data~\cite{little02}. A variety of statistical models has been developed for missing data under MNAR. For example, \citeA{diggle94} proposed a general framework based on selection modeling approach, while \citeA{little93} adopted a pattern mixture modeling approach. These models in general are highly dependent on unidentified assumptions~\cite{little95,kenward98}. As a result, statisticians have been developing the sensitivity analysis, where unidentified parameters are varied over a plausible range~\shortcite{little94,rotnitzky98,schar99,daniels00,verbeke01,kenward01,thijs02}.

Although advocating sensitivity analysis is a very useful strategy to evaluate the effect of departures from unidentified assumptions, it might be dissatisfactory for some researchers since no single summary is provided. In this context, Bayesian approaches are appealing as they allow to formally incorporate a priori knowledge about the unidentified parameters. For pattern mixture models, the early publication of~\citeA{rubin77} has provided a general framework to express the assumptions about the differences between respondents and nonrespondents as prior distributions. For selection models, a series of semiparametric methods has developed by~\shortciteA{schar99,schar03,rotnitzky01} that incorporate prior beliefs about the selection mechanism. A general overview of methods is provided in~\citeA{daniels08}.

This paper presents a new approach that allows the researchers to draw a `single' conclusion using the observed data. We build our approach on the basis of the proposed model by~\shortciteA{rotnitzky01} or \shortciteA{schar03}, where strong assumptions are placed on missingness mechanism as opposed to the distributional form of the outcome.

A starting point for analysis is to assume that the missingness mechanism is MAR.  to evaluate the effect of departures from these assumptions


Multiple Imputation (MI) is an accepted and effective way to deal with missing data~\cite<see, for a general review,>{kenward07}. Given the complete data, MI assumes that the distribution of the missing part of the incomplete variable is the same as the distribution of the observed part of the incomplete variable, provided that the missingness mechanism is MAR. Suppose we have an incomplete variable $X$ and a fully observed variable $Y$. Let $X_{obs}$ be the observed part of $X$, and $X_{mis}$ be the missing part. Given $Y$, MI uses $X_{obs}$ to determine the distribution from which imputations will be drawn, and makes use of this distribution to draw imputation values. This distribution, however, is not appropriate for drawing imputations under MNAR. Distributions of $X_{obs}$ and $X_{mis}$ may have different locations or dispersions under MNAR, and such differences need to be taken into account in creating imputations. As the values of $X$, by definition, do not contain appropriate information to estimate such differences, the amount of these differences are not known. The state of the art is to choose different adjustment values, i.e. to conduct a sensitivity analysis and study the effect of different choices of the adjustment. Under MNAR, the values of $X$ should be a part of the missingness mechanism but, since $X$ is partly missing, it is difficult to create model for the missingness mechanism that includes $X$.

In this paper, we attempt to find a way out of this dilemma by iteratively imputing the incomplete variable $X$ and remodeling the missingness mechanism of $X$. We restrict our attention to distributions that change only in location as a consequence of nonresponse. Specifically, we assume that the observed and missing parts of $X$ have similar variances but different means. To develop this idea, it is useful to define a response indicator $R$ that represents whether or not a measurement is observed. The key point is to generate a pseudo response indicator $\dot{R}$ by drawing random values from the model for the missingness mechanism. As we will show, this additional variable in combination with $R$ provides important clues about the amount of the adjustment needed. 

The idea of drawing $\dot{R}$ is called the \textit{Random Indicator} (RI) method for imputing data under MNAR. An iterative algorithm successively generates the pseudo response indicator $\dot{R}$ and imputations $\dot{X}$ for the missing values. The RI method is automatic and provides a new approach to the data that are MNAR. 

This article is built up as follows. In the next section, we define the models for the data that are MNAR. Section 3 proposes the RI method in details. Section 4 reports a comprehensive simulation study to evaluate the performance of the proposed imputation method. Section 5 analyses our real life data. Concluding remarks are given in the last section.

\section{Model}

We consider the case of univariate missing data in order to illustrate the imputation process for nonignorable nonresponse. Let $\X = (X_{1}, \ldots, X_{k})\p$ denote a $k-$dimensional vector of variables, where $X_1$ is subject to missing data and $X_{2}, \ldots, X_{k}$ are fully observed. Here, we do not distinguish between the outcome and explanatory variables since any of them can be subject to missing data. Our interest lies in estimating some parameter vector $\b$, e.g. coefficients in the Cox regression or linear regression model. Since the parameter vector $\b$ may not necessarily completely specify the distribution of $\X$, we index the probability density function of $\X$ by $\t$, i.e. $f(\X;\t)$.


Suppose that the missingness of $X$ depends on two types of variables: the incomplete variable $X$ and the complete variable $Y$. These make the missingness mechanism MNAR. Let $R$ denote a response indicator such that $R=1$ if $X$ is observed, and $R=0$ if it is missing. Let the logit of the probability for $X$ to be observed be as follows:
\begin{eqnarray}
\label{nonres}
{\rm{logit}}\{P(R = 1|X, Y; \psi)\} = \psi_0 + \psi_1X + \psi_2Y,
\end{eqnarray}
where $\psi = (\psi_0, \psi_1, \psi_2)\p$ are unknown parameters in the so-called nonresponse model. Note that formulation~(\ref{nonres}) is sufficiently general to allow the dependence of the probability of response on all information. If we set $\psi_1 = 0$ in model~(\ref{nonres}), the missingness mechanism is MAR. Reversely, if we set $\psi_2 = 0$ in model~(\ref{nonres}), the nonresponse model is the most extreme form of MNAR as the missingness depends on the incomplete variable $X$ only.

Distributions of the observed part of $X_1$ and the missing part of it are not necessarily the same under MNAR. Define the observed part of $X_1$ by $X_{obs}$, and the missing part by $X_{mis}$, so that $X_1 = (X_{obs}, X_{mis})$. As already mentioned in the introduction, we consider a situation where $X_{obs}$ and $X_{mis}$ have equal variances but unequal means. This allows missingness to affect the location of the distribution of $X_1$ only. The following proposition shows how the distribution of $X_{mis}$ relates to the distribution of $X_{obs}$.
\begin{proposition}
\label{proposi1}
Let the logit of the probability for $X_1$ to be observed be as model~(\ref{nonres}). If $X_{obs}$ conditional on $\Z$ is normally distributed with mean $\mu_{z}$ and variance $\s$, then
$$X_{mis}\sim N(\mu_{z} - \d, \s),$$
where $\d = \psi_1\s$.
\end{proposition}
Proposition~\ref{proposi1} implies that if we assume a logistic function for the probability for $X_1$ to be observed as model~(\ref{nonres}), the normality of $X_{obs}$ results in the normality of $X_{mis}$ where the variances are the same, but the means are not (see Appendix A.1 for proof). 

The difference between the means of $X_{obs}$ and $X_{mis}$, i.e. $\d$, plays an important role in the imputation of missing data. Under MAR, $\psi_1$ is equal to zero in model~(\ref{nonres}), and as a result $\d = 0$. Different from MAR, we are interested in MNAR cases where $\d\neq 0$. Because $\d$ is unknown, the current advice is to pre-choose different values for $\d$ over a reasonable range, and study the effect of pre-chosen values on the inferences~\shortcite{buuren99,carpenter07,pfeff11}. As opposed to the current advice, the RI method, as described in the next section, attempts to estimate $\d$ from the observed data.

It is worth mentioning that this way of modeling MNAR data is similar to pattern-mixture models~\cite{little93} where the distribution of $X_1$ is characterized conditional on the response indicator $R$. But, it is different from selection models~\cite{diggle94} where the marginal distribution of $X_1$ is assumed to be normal.

\section{Random Indicator Imputation}
\subsection{$\dot{R}$ given}
To motivate our method under nonignorable nonresponse, we consider a situation where an independent realization of the response indicator $R$ is available. This is to say that another realization of $R$, apart from $R$, is independently generated from the nonresponse model~(\ref{nonres}). This realization is called pseudo indicator $\dot{R}$.

Note that $R$ and $\dot{R}$ partition the incomplete variable $X_1$ into four subsets. The first subset, which we call the reference group, consists of measurements where $R=1$ and $\dot{R}=1$, the second subset consists of measurements where $R=1$ and $\dot{R}=0$, and so on. Let us define four means
\begin{eqnarray*}
\mu_{11} &=& E(X_1 | R = 1, \dot{R} = 1)\\
\mu_{10} &=& E(X_1 | R = 1, \dot{R} = 0)\\
\mu_{01} &=& E(X_1 | R = 0, \dot{R} = 1)\\
\mu_{00} &=& E(X_1 | R = 0, \dot{R} = 0),
\end{eqnarray*}
where $\mu_{11}$ is the marginal expectation of $X_1$ given $R = 1$ and $\dot{R} = 1$ (the reference group), and so on. Furthermore, we define
\begin{eqnarray*}
\d_{R}  &=& \mu_{11} - \mu_{10}\\
\d_{NR} &=& \mu_{01} - \mu_{00},
\end{eqnarray*}
where $\d_{R}$ is the difference between the means of $X_1$ when $(R = 1, \dot{R} = 1)$ and $(R = 1, \dot{R} = 0)$ in the observed data, and $\d_{NR}$ is the difference of $X_1$'s means when $(R = 0, \dot{R} = 1)$ and $(R = 0, \dot{R} = 0)$ in the missing data. For a given $\dot{R}$, $\d_{R}$ can be calculated from the observed part of $X_1$ while $\d_{NR}$ depends on the missing part of $X_1$ and cannot be calculated.

Since both $R$ and $\dot{R}$ are two realizations of the same nonresponse model, it holds by definition that $P(R = 1) = P(\dot{R} = 1)$. If the realizations are independent, it is not difficult to prove that the means of $X_1$ when either $(R = 1, \dot{R} = 0)$ or $(R = 0, \dot{R} = 1)$ are identical (the proof is given in Appendix A.2):

\begin{eqnarray*}
\label{f1}
\mu_{10} = \mu_{01}.
\end{eqnarray*}
\begin{figure}[h]
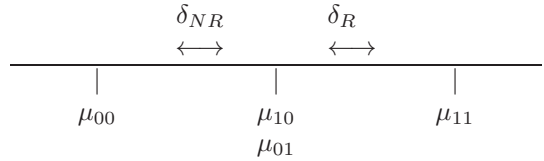

\centering
{\footnotesize
\begin{tabular}{p{0.25cm}cp{1cm}cp{1cm}cp{0.25cm}}
            &\multicolumn{2}{r}{$\d_{NR}$} && \multicolumn{2}{l}{$\d_{R}$} & \\
            &\multicolumn{2}{r}{$\longleftrightarrow$} && \multicolumn{2}{l}{$\longleftrightarrow$} & \\
\hline
            &$\mid$      && $\mid$     && $\mid$       \\
            &$\mu_{00}$  && $\mu_{10}$ && $\mu_{11}$   \\
            &            && $\mu_{01}$ &&              \\
\end{tabular}
\caption{Graphical representation of the means and their differences}
\label{fig}
}
\end{figure}

To clarify the notation, we represent the four means and their differences in Figure~\ref{fig} in the case that more missing data occur in the lower part of the distribution of $X_1$. As can be seen, $\mu_{10}$ equals $\mu_{01}$, and $\mu_{11}$ and $\mu_{00}$ are located in the same distance from the middle, i.e. $\mu_{10}$. It implies that the middle serves as a pivot for $\mu_{11}$ and $\mu_{00}$. The question is how $\d_{R}$ and $\d_{NR}$ are related. It can be shown that $\d_{R}$ and $\d_{NR}$ are identical in certain circumstances. Proposition~\ref{proposi2} provides a justification for equality of $\d_{R}$ and $\d_{NR}$. 
\begin{proposition}
\label{proposi2}
Let $R$ and $\dot{R}$ be two independent draws from the nonresponse model~(\ref{nonres}). If $X_1$ given $\Z$, $R = 1$ and $\dot{R} = 1$ is normally distributed with mean $\mu_z$ and variance $\s$, then
$$X_1|\Z, R = r, \dot{R} = \dot{r} \sim N(\mu_z + \psi_1(r + \dot{r} - 2)\s, \s) \qquad r, \dot{r} = 0, 1,$$
and as a result
\begin{eqnarray*}
\d_R    &=& E[X_1|R=1, \dot{R} = 1] - E[X_1|R=1, \dot{R} = 0] = \psi_1\s\\
\d_{NR} &=& E[X_1|R=0, \dot{R} = 1] - E[X_1|R=0, \dot{R} = 0] = \psi_1\s.
\end{eqnarray*}
Hence,
$$\d_R=\d_{NR}.$$
\end{proposition}

Proposition~\ref{proposi2} implies that if we assume normality for $X_1$ given $\Z$, $R = 1$ and $\dot{R} = 1$, we find $\d_{R}=\d_{NR}$ (the proof is a direct extension of Proposition~\ref{proposi1}). 

The idea is to calculate $\d_R$ from the observed part of the data, and use it as an adjustment to the missing part. Hence, one can impute the unobserved data under the MAR assumption, and then correct for the imputations by utilizing the amount of the estimated $\d_R$ under the specified MNAR mechanism. More specifically, we can impute missing data as follows. Let us define $\d_{adj} = \d_{R} = \d_{NR}$ for notation convenience. The amount of the adjustment, i.e. $\d_{adj}$, can be estimated from the observed part of the data in the following equation:
\begin{eqnarray}
\label{f2}
E(X_1|\Z, R = 1, \dot{R} = \dot{r}) &=& \Z\p\phi + \d_{adj}(\dot{r} - 1),
\end{eqnarray}
and then missing data are imputed using the imputation model
\begin{eqnarray}
\label{f3}
E(X_1|\Z, R = 0, \dot{R} = \dot{r}) &=& \Z\p\phi + \d_{adj}(\dot{r} - 2),
\end{eqnarray}
for $\dot{r} = 0, 1$ where $\phi$ (which is some function of $\t$) is the parameters corresponding to complete covariate $\Z$ in the imputation model. The method as developed thus far consists of a few simple steps that are summarized in the following algorithm: 

\begin{algo}
Imputation of $X_{mis}$ for a given $\dot{R}$
\end{algo}
\begin{enumerate}
  \item Calculate $\hat\phi$ and $\hat\d_{adj}$ from the observed part of the data by equation~(\ref{f2}).
  \item Draw $\dot\phi$ from its posterior distribution for a given prior for $\phi$.
  \item Predict the missing data for the part $(R = 0, \dot{R} = 1)$ using $\Z\dot\phi - \hat\d_{adj}$.
  \item Predict the missing data for the part $(R = 0, \dot{R} = 0)$ using $\Z\dot\phi - 2\hat\d_{adj}$.
  \item Impute the missing data by adding an appropriate amount of noise to the predicted values in steps 3 and 4 as usual.
\end{enumerate}

\subsection{$\dot{R}$ drawn}
In practice, the missing part of $X_1$ is unknown. Consequently, the probability of the response given $X_1$ in model~(\ref{nonres}) is also unknown. Thus, in general, we cannot generate the pseudo indicator variable $\dot R$. However, we can generate $\dot R$ in an algorithm that iterates over equations~(\ref{nonres}),~(\ref{f2}) and~(\ref{f3}). 

To develop this idea in the imputation framework, we start with the posterior predictive distribution $P(X_{mis}|X_{obs}, R)$. The notation is simplified by suppressing the fully observed covariates $\Z$. Embedding this posterior predictive distribution into a model on a larger space such as below might be useful. This is a key principle in computational methods. The posterior predictive distribution $P(X_{mis}|X_{obs}, R)$ can thus be written as:
\begin{eqnarray*}
P(X_{mis}|X_{obs}, R) &=& \int\int P(X_{mis}, \dot{R}|X_{obs}, R, \gamma)P(\gamma|X_{obs}, R)\partial\gamma \partial\dot{R},
\end{eqnarray*}
where $\gamma = (\t, \psi)$. The above equation implies that the posterior predictive distribution $P(X_{mis}|X_{obs}, R)$ can be simulated by drawing a value of $\gamma$ from its posterior distribution, $P(\gamma|Y_{obs}, R)$, and then a value of $X_{mis}$ and a pseudo response indicator $\dot{R}$ are drawn from their joint conditional distribution, $P(X_{mis}, \dot{R}|X_{obs}, R, \gamma)$. 

Drawing a sample from the joint conditional distribution $P(X_{mis}, \dot{R}|X_{obs}, R)$ is a challenging task. In the spirit of the Gibbs sampler, we use an iterative algorithm which runs between the following two steps:
\begin{eqnarray*}
X_{mis} &\sim& P(X_{mis}|X_{obs}, R, \dot{R})\\
\dot{R} &\sim& P(\dot{R}|X_{obs}, R, X_{mis}).
\end{eqnarray*}
The procedure is motivated by the fact that drawing from two univariate conditional distributions is easier than drawing from a joint distribution. Using univariate conditional distributions is quite popular in practice when drawing from the joint distribution is difficult~\shortcite{white11,buuren06,ragu01}. Here, we implicitly assume that draws from the joint distribution are generated by successively drawing from each univariate conditional distribution. Although there is no guarantee for the existence of the joint distribution from which the values are drawn, experience has shown that it often leads to valid statistical inferences in a variety of cases~\cite{buuren07}.

In this stage , we describe how $\dot R$ and $X_{mis}$ can be drawn iteratively. We first start with the initially completed-data $\dot X_1^{0} = (X_{obs}, \dot X^{0}_{mis})$ where $\dot X^{0}_{mis}$ is a random draw from $X_{obs}$. The parameters in the nonresponse model~(\ref{nonres}) are estimated given $\dot X^{0}$. We then draw random values from the posterior distributions of the parameters in the nonresponse model~(\ref{nonres}), i.e. $\dot\psi$, and calculate the probability of response given $\dot X_1^{0}$. The initial pseudo response indicator $\dot R$ is generated by a standard procedure for generating a binary variable. The drawn values for $\dot R$ are used to estimate $\d_{adj}$ by equation~(\ref{f2}), and the imputed values of the missing data are updated by equation~(\ref{f3}). These sequential steps are repeated consistently for a sufficient number of iterations. Finally, we summarize the whole procedure, i.e. the RI method, in the following algorithm:
\begin{algo}
Imputation of $X_{mis}$ in general
\end{algo}
\begin{enumerate}
  \item Draw initial values $\dot X^{0}_{mis}$ randomly from $X_{obs}$.
  \item Draw $\dot\psi$ from its posterior distribution for a given prior for $\psi$.
  \item Draw $\dot R$ from a Bernouli process given $\psi = \dot\psi$
  \item Impute $\dot X_{mis}$ by algorithm (1).
  \item Return to step 2 to iterate the algorithm (2) for a few number of times.
\end{enumerate}

It should be mentioned that this algorithm usually needs a few iterations. It is common to iterate such an algorithm for a small number of times in practice although there are a number of approaches to check the convergency of the algorithm \shortcite{Adlouni06}. After iterating for a small number of times, say 10 or 20, the last drawn values are treated as the imputed values under MNAR. At this point, multiple imputations can be generated by starting from different $\dot X^{0}_{mis}$.




\section{Simulation Study}
The RI method is evaluated by a comprehensive simulation study. For this purpose, we define two versions of the complete data model and five versions of the nonresponse model. We also consider two different sample sizes. The quality of the method is evaluated in terms of bias and $95\%$ coverage rate of the estimated parameters of primary interest. Moreover, we perform complete case analysis and the traditional multiple imputation to investigate the effect of the underlying assumption of nonignorability.

\subsection{Set-up}
The simulation starts by defining a model of interest which is a linear regression of $X_1$ on covariates $X_2$ and $X_3$. We define the model of interest as $X_1 = \b_1 + \b_2X_2 + \b_3X_3 + \varepsilon$ where $\b = (\b_1, \b_2, \b_3)$ are regression coefficients in which we are interested. Covariates $X_2$ and $X_3$ and the error term $\varepsilon$ are generated from independent normal random variables with the following specifications: $X_2\sim N(2, 4)$, $X_3\sim N(-1, 1)$, and $\varepsilon\sim N(0,1)$. We consider two sets of coefficients in order to have a data set with strong and moderate associations between $X_1$ and $(X_2, X_3)$ (in terms of explained variance, $R^2$). For instance, $\b = (1, 0.5, 1)$ results in, on average, 65 percent of the association ($R^2 = 0.65$) whereas a weaker association, about 32 percent ($R^2 = 0.32$), can be achieved by a set of $\b = (3, -0.25, 0.5)$.

The probability of response, denoted by $p$, is measured by a logistic function. This function can take the form
$${\rm{logit}}(p) = \psi_0 + \psi_1X_1 + \psi_2Z$$
where $Z = X_2$. The response indicator $R$ is then generated from a Bernouli process that corresponds to $p$. It is worth mentioning that, by utilizing the response indicator $R$, the simulated complete variable $X_1$ is partitioned into the observed part and the missing part, and the mean of the missing part differs from the mean of the observed part. Similar to the regression coefficients in the previous step, five scenarios are considered for the coefficients in the nonresponse model, i.e. $\psi = (\psi_0, \psi_1, \psi_2)$. In total, there are two complete data models that are combined with five nonresponse models. Table~\ref{set} shows the missing data percentages for the setup values for the simulation study.
\begin{table}
\centering
{\footnotesize
\caption{Setup values and the missing data percentages}
\label{set}
\smallskip
\begin{tabular}{llcc}
  \hline\hline
  &&\multicolumn{2}{c}{$\b$}\\
  \cline{3-4}mechanism &\multicolumn{1}{c}{$\psi$} & $(1, 0.5, 1)$ & $(3, -0.25, 0.5)$\\
  \hline
  MCAR  & $(-0.75, 0.00, 0.00)$ & 68 & 68  \\
  MAR   & $(-2.00, 0.00, 0.50)$ & 70 & 70  \\
  MNAR1 & $(-0.50, 0.50, 0.25)$ & 41 & 28  \\
  MNAR2 & $(-1.00, 0.75,-0.50)$ & 73 & 58  \\
  MNAR3 & $(-2.00, 1.50, 0.00)$ & 57 & 35  \\
  \hline\hline
\end{tabular}
}
\end{table}

Different parameter choices correspond to different missingness rates and different missingness mechanisms. The first scenario where $\psi=(-0.75, 0.0, 0.0)$ gives a constant probability of response to all observations, i.e. MCAR. The second scenario gives the MAR mechanism where the response probabilities does not depend on $X_1$. The other three scenarios give different MNAR mechanisms. MNAR1 and MNAR2 are moderate versions of nonignorable missing data in the sense that the response probabilities depend on both completely observed variable $Z$ and incomplete variable $X_1$. The last scenario, i.e. MNAR3, describes the most extreme case of MNAR where the response probabilities depend on $X_1$ only.

Random samples of size $n = 200$ and $n = 1000$ were taken to investigate the small and large sample properties of our proposed method. The number of replications was 1000, and the number of the Gibbs sampler iterations was set to 10. The number of imputed data sets was 5. All calculations were done in R~11.2.1.

The overall mean of the estimated regression coefficient for $\b_1$, $\b_2$ and $\b_3$ as well as the $95\%$ coverage rate was calculated. We compared our method (RI) with the complete case (CC) method where all missing data were ignored. We also added the multiple imputation (MI) method under MAR to the analysis. For this purpose, we used MICE (MICE, v 2.8; \citeA{buuren11}).

\subsection{Result}
Table~\ref{nlarge} shows the results of the simulation study in the large sample ($n = 1000$).
\begin{table}[!h]
\centering
{\scriptsize
\caption{Estimates of the regression coefficients $\b$ and coverage probabilities of $95\%$ (in parentheses) for sample size $n = 1000$.}
\label{nlarge}
\smallskip
\begin{tabular}{lcccccccc}
\hline\hline
\multicolumn{2}{l}{Mechanism}& \multicolumn{3}{c}{Strong Association}  && \multicolumn{3}{c}{Moderate Association}\\
\cline{3-5}\cline{7-9} &     & $\b_1$ & $\b_2$ & $\b_3$ && $\b_1$ & $\b_2$ & $\b_3$ \\
                       &True & 1.000  & 0.500  & 1.000  && 3.000  & -0.250 & 0.500 \\
\hline
     & CC  & 1.001(95) & 0.500(97) & 1.003(95) && 3.000(94) & -0.250(96) & 0.499(96) \\
MCAR & MI  & 1.003(94) & 0.501(94) & 1.004(93) && 3.003(94) & -0.249(94) & 0.504(93) \\
\smallskip
     & RI  & 1.004(95) & 0.500(92) & 1.004(92) && 3.005(96) & -0.250(93) & 0.504(91) \\
     & CC  & 0.998(95) & 0.500(96) & 0.998(95) && 2.998(95) & -0.250(96) & 0.498(95) \\
MAR  & MI  & 0.998(93) & 0.500(95) & 0.998(94) && 2.997(93) & -0.250(95) & 0.498(94) \\
\smallskip
     & RI  & 0.998(95) & 0.500(95) & 0.997(92) && 2.989(97) & -0.249(95) & 0.497(92) \\
     & CC  & 1.230(17) & 0.458(55) & 0.958(84) && 3.137(44) & -0.262(91) & 0.478(91) \\
MNAR1& MI  & 1.230(26) & 0.458(63) & 0.958(86) && 3.137(47) & -0.262(90) & 0.477(88) \\
\smallskip
     & RI  & 0.993(94) & 0.504(93) & 1.004(96) && 3.021(95) & -0.251(95) & 0.500(95) \\
     & CC  & 1.370(01) & 0.518(92) & 0.899(68) && 3.191(23) & -0.167(15) & 0.457(88) \\
MNAR2& MI  & 1.370(05) & 0.517(92) & 0.899(72) && 3.192(26) & -0.167(29) & 0.457(87) \\
\smallskip
     & RI  & 0.971(95) & 0.506(92) & 0.964(90) && 3.039(95) & -0.225(90) & 0.489(94) \\
     & CC  & 1.617(00) & 0.390(01) & 0.778(00) && 3.155(30) & -0.196(22) & 0.392(20) \\
MNAR3& MI  & 1.620(00) & 0.390(04) & 0.777(04) && 3.154(26) & -0.196(21) & 0.391(22) \\
     & RI  & 1.066(87) & 0.481(89) & 0.959(88) && 3.049(86) & -0.251(96) & 0.501(95) \\
\hline\hline
\end{tabular}
\vspace{-.25cm}
\begin{flushleft}
\emph{Note:} Given are complete case analysis (CC), multiple imputation under MAR (MI), and multiple imputation under MNAR (RI).
\end{flushleft}
}
\end{table}
In the first and second scenarios, all methods worked alike, and the parameters and coverage rates were accurately estimated. Apparently, in the second scenario where MAR holds, CC and MI were equal. This is not surprising because, here, we have a special situation where missing data occur in the outcome variable $X_1$ only, and the complete data model is correct. Thus, CC and MI are equivalent here, and produce the same results (see,~\cite{buuren18} and references on it). In the MNAR scenarios, CC and MI (which assume ignorability) went far off. In particular, when the association was strong, and the scenario MNAR3 holds, the coverage probabilities dramatically dropped to zero. The results of CC and MI were similar, but coverage of MI was often better. In contrast, RI provided essentially correct results in all cases. The relative bias, $(\bar{\hat\b} - \b)/\b$, was less than $8\%$ of the true value for the regression coefficients, and the coverage rates were in an acceptable range. In rare cases, we saw coverage rates around $88\%$ especially in strong association, which is presumably due to the extreme probability of nonresponse.

\begin{table}[!h]
\centering
{\scriptsize
\caption{Estimates of the regression coefficients $\b$ and coverage probabilities of $95\%$ (in parentheses) for sample size $n = 200$.}
\label{nsmall}
\smallskip
\begin{tabular}{lcccccccc}
\hline\hline
\multicolumn{2}{l}{Mechanism}  & \multicolumn{3}{c}{Strong Association}  && \multicolumn{3}{c}{Moderate Association}\\
\cline{3-5}\cline{7-9} &     & $\b_1$ & $\b_2$ & $\b_3$ && $\b_1$ & $\b_2$ & $\b_3$ \\
                       &True & 1.000  & 0.500  & 1.000  &&  3.000 & -0.250 & 0.500  \\
\hline
     & CC  & 1.016(96) & 0.498(95) & 1.005(97) && 2.999(95) & -0.250(95) & 0.503(95) \\
MCAR & MI  & 1.018(95) & 0.497(93) & 1.005(96) && 3.002(94) & -0.250(94) & 0.504(95) \\
\smallskip
     & RI  & 1.023(95) & 0.500(93) & 1.010(94) && 2.997(97) & -0.253(93) & 0.504(93) \\
     & CC  & 1.000(95) & 0.504(96) & 1.008(95) && 3.000(95) & -0.246(96) & 0.508(95) \\
MAR  & MI  & 0.995(94) & 0.505(94) & 1.008(92) && 2.995(94) & -0.245(94) & 0.508(92) \\
\smallskip
     & RI  & 1.003(96) & 0.502(94) & 1.011(91) && 2.973(96) & -0.245(94) & 0.509(91) \\
     & CC  & 1.230(76) & 0.459(89) & 0.961(95) && 3.137(86) & -0.261(96) & 0.478(96) \\
MNAR1& MI  & 1.228(78) & 0.459(88) & 0.962(92) && 3.136(86) & -0.260(94) & 0.479(95) \\
\smallskip
     & RI  & 0.998(95) & 0.505(95) & 1.013(94) && 3.032(96) & -0.251(94) & 0.509(95) \\
     & CC  & 1.371(59) & 0.518(98) & 0.906(93) && 3.186(81) & -0.168(78) & 0.458(95) \\
MNAR2& MI  & 1.369(65) & 0.519(94) & 0.906(91) && 3.186(79) & -0.168(79) & 0.458(93) \\
\smallskip
     & RI  & 0.982(95) & 0.507(94) & 0.973(93) && 3.056(95) & -0.226(94) & 0.492(94) \\
     & CC  & 1.621(14) & 0.390(58) & 0.774(57) && 3.155(83) & -0.195(80) & 0.391(81) \\
MNAR3& MI  & 1.622(20) & 0.390(59) & 0.772(60) && 3.154(77) & -0.195(75) & 0.391(74) \\
     & RI  & 1.077(92) & 0.478(93) & 0.952(94) && 3.058(94) & -0.250(94) & 0.502(94) \\
\hline\hline
\end{tabular}
\vspace{-.25cm}
\begin{flushleft}
\emph{Note:} Given are complete case analysis (CC), multiple imputation under MAR (MI), and multiple imputation under MNAR (RI).
\end{flushleft}
}
\end{table}

The simulation results in small sample ($n = 200$) are illustrated in Table~\ref{nsmall}. Overall, the results are similar to those in Table~\ref{nlarge}. All methods performed equally under MCAR and MAR. In different cases of MNAR, CC and MI were not as accurate as RI, in particular when the nonresponse probabilities were in the most extreme scenario MNAR3. Large relative biases were found in the estimates of the intercept $\b_0$ in both CC and MI, and also the coverage rates were low. In contrast, RI performed successfully. Estimates of the parameters were essentially unbiased with high coverage rates. The relative bias was not more than $8\%$.

\begin{figure}[!h]
\centering
{\footnotesize
\includegraphics[scale=.80, trim = 10mm 55mm 0mm 55mm]{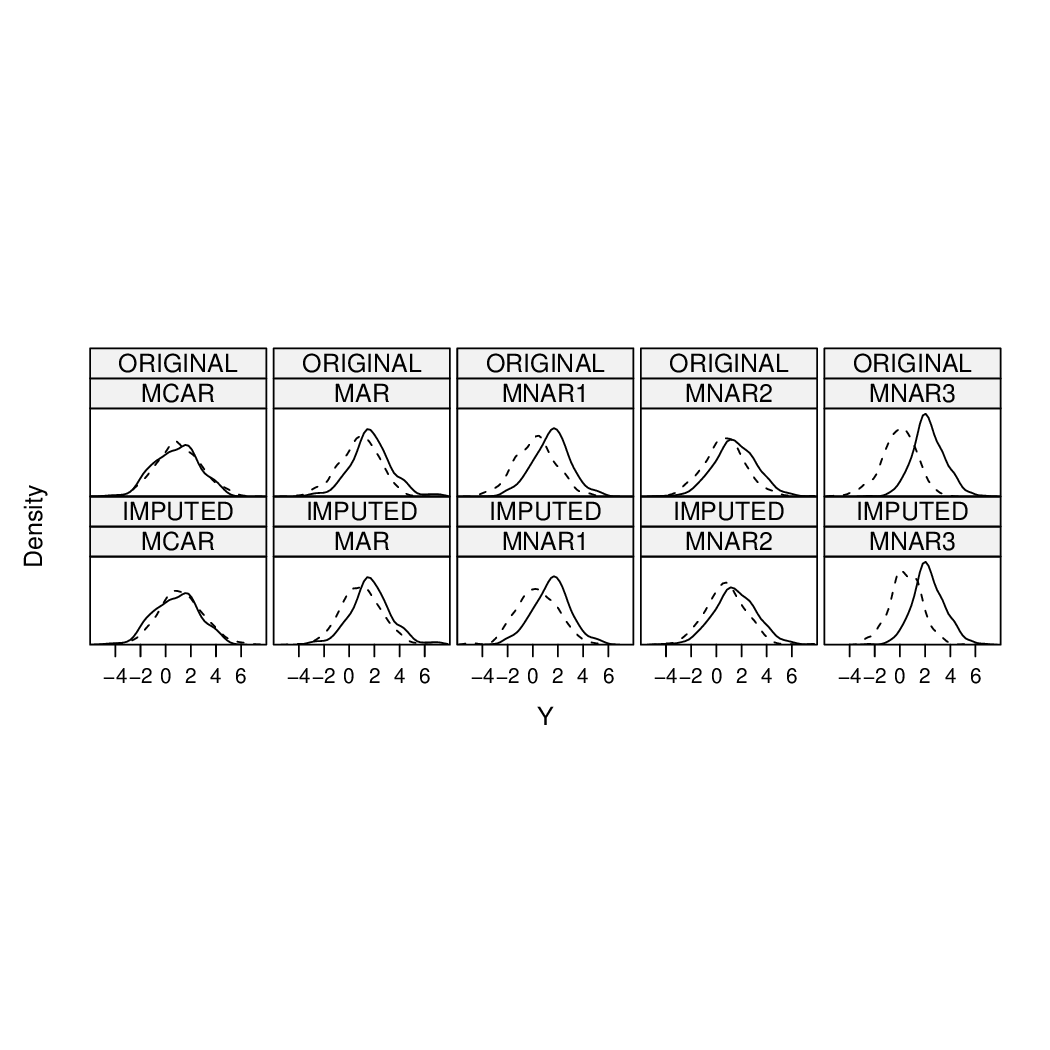}
\caption{Density plot of the original (top) and imputed (bottom) data. The solid line refers to the observed data, and the dashed line refers to the original values before creating missingness (top) and after imputing them (bottom) for different scenarios by the RI method.}
\label{obsmis}
}
\end{figure}

The appropriateness of the RI method was also investigated by comparing the original and imputed data. Figure~\ref{obsmis} illustrates kernel density estimates for the original data (top) and the imputed data (bottom). The solid line refers to the density of the observed data, which is essentially the same in the original and imputed data. The dashed line refers to the density of the values before creating missing data (top) and the density of the imputed data (bottom). In general, the dashed lines in the bottom line are close to the top line. This implies that the RI method imputes missing values pretty close to the original ones. Under MCAR, the observed and missing parts of the original data were distributed identically (top left). Likewise, the imputed and observed data were similar in the completed-data (bottom left). In other scenarios and for the original data, we observed that the missing part of the data are separated from the observed part of the data. The same holds for the completed-data (after imputation). The last scenario is the most extreme case where they are mainly separated (top right). The same pattern also observes for the completed-data (bottom right).

\begin{figure}[h]
\centering
{\footnotesize
\includegraphics[scale=.70, trim = 10mm 10mm 10mm 110mm]{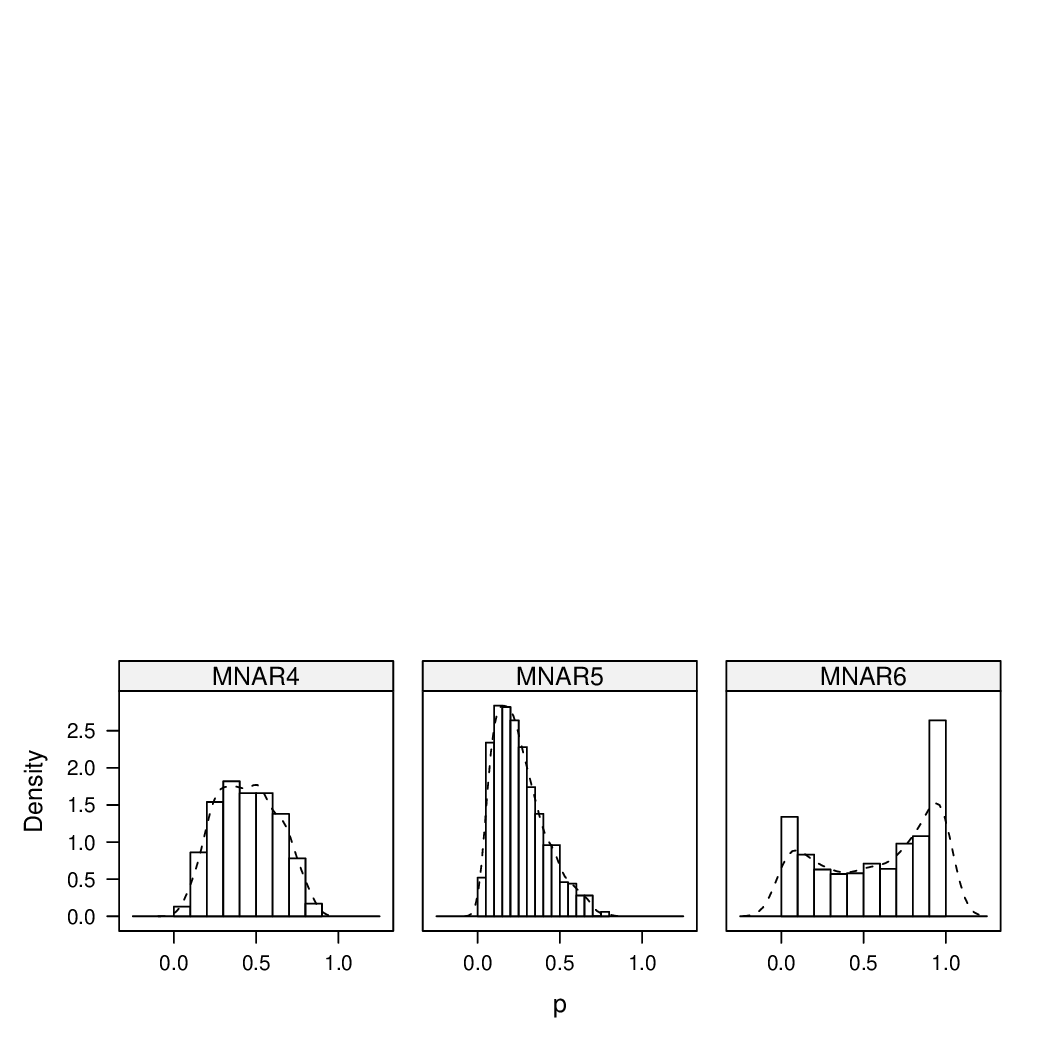}
\caption{Histogram of the probability of being observed with a kernel density estimate}
\label{prob}
}
\end{figure}

To more closely investigate the performance of the RI method, we further considered the most extreme case of MNAR where the response probabilities depend on the incomplete variable $X_1$ only. Three different conditions of the MNAR mechanism were considered to produce different patterns of missingness. These conditions represent situations where the response probabilities are symmetrically distributed (MNAR4), have skewed pattern (MNAR5), or are more likely in both tails (MNAR6). Figure~\ref{prob} shows histograms of the observed probability ($p$) for scenarios MNAR4 - MNAR6. By these, we can test the RI method under different patterns of $p$ when the missingness mechanism is the most extreme case of MNAR. 

Table~\ref{ext} shows only the simulation results for the large sample size because of space limitation. As can be seen, CC and MI gave biased results in all conditions, and coverage rates marginally declined, particularly for the estimate of $\t_1$. Interestingly, under the strong association, coverage rates rapidly fell to zero in MNAR6. The RI method, on the other hand, performed very well. Biases were negligible, and coverage rates achieved the nominal level.
\begin{table}[!h]
\centering
{\scriptsize
\caption{Estimates of the regression coefficients $\b$ and coverage probabilities of $95\%$ (in parentheses) for the most extreme nonresponse model for sample size $n = 1000$.}
\label{ext}
\smallskip
\begin{tabular}{lcccccccc}
\hline\hline
\multicolumn{2}{l}{Mechanism}& \multicolumn{3}{c}{Strong Association}  && \multicolumn{3}{c}{Moderate Association}\\
\cline{3-5}\cline{7-9} &     & $\b_1$ & $\b_2$ & $\b_3$ && $\b_1$ & $\b_2$ & $\b_3$ \\
                       &True & 1.000  & 0.500  & 1.000  && 3.000  & -0.250 & 0.250 \\
\hline
     & CC  & 1.258(14) & 0.475(82) & 0.949(83) && 3.159(35) & -0.238(92) & 0.473(91) \\
MNAR4& MI  & 1.263(23) & 0.474(84) & 0.949(86) && 3.156(43) & -0.236(90) & 0.474(91) \\
\smallskip
     & RI  & 0.996(95) & 0.499(96) & 0.998(93) && 3.008(96) & -0.249(94) & 0.501(94) \\
     & CC  & 1.361(11) & 0.477(89) & 0.955(88) && 3.265(10) & -0.237(94) & 0.473(94) \\
MNAR5& MI  & 1.363(28) & 0.476(90) & 0.955(90) && 3.264(25) & -0.237(92) & 0.475(93) \\
\smallskip
     & RI  & 1.014(95) & 0.493(91) & 0.988(91) && 3.003(94) & -0.247(92) & 0.496(94) \\
     & CC  & 1.426(00) & 0.409(01) & 0.816(01) && 3.051(87) & -0.211(40) & 0.420(39) \\
MNAR6& MI  & 1.424(00) & 0.408(03) & 0.815(03) && 3.055(83) & -0.212(38) & 0.424(41) \\
     & RI  & 1.025(94) & 0.505(94) & 1.009(95) && 3.041(90) & -0.255(94) & 0.510(96) \\
\hline\hline
\end{tabular}
\vspace{-.25cm}
\begin{flushleft}
\emph{Note:} Given are complete case method (CC), multiple imputation under MAR (MI), and multiple imputation under MNAR (RI).
\end{flushleft}
}
\end{table}

We have seen that, based on the simulation results, the RI method performed very well in all situations. To summarize, it appears that the RI method performs better than the other methods, both with and without covariate information.


\section{Application}
\subsection{The Leiden $85^+$ cohort}
We analyse the blood pressure (BP) data that have previously been studied in~\shortciteA{buuren99}. It has been found that the mortality of elderly
people aged 85 and older is increased among those with lower BP~\cite{glynn95}. The idea of the cohort study was to determine whether this
inverse association between BP and mortality could be explained by the health measurements in this age group. The number of people in the cohort
was 1236 of which 956 were interviewed. A detailed description of the cohort study was reported elsewhere~\shortcite{lag91, lag92}.

The relation between mortality and BP is investigated by a Cox regression model adjusted for sex, age, and health measurements such as hypertension, haemoglobin, and serum cholesterol. This analysis is confronted by missing data. About 12.5 per cent of the individuals had missing observations for the BP. It was suspected that the missing values had happened among individuals with lower BP \shortcite{buuren99}. For instance, if the respondent used any anti-hypertension drugs, the BP was measured less frequently. \shortciteA{buuren99} argued that the unobserved BP might be MNAR and conducted a simple adaptation method to adjust for the imputations that were created under MAR. Then, the sensitivity of the inferences was investigated against violations from MAR by choosing different adjustment values. We here use the RI method to handle this problem.

The BP was measured by systolic and diastolic pressure. We use the systolic BP (SBP) for the analysis only. Our strategy is to impute the missing values in the SBP using the newly developed imputation method, and then investigate the relation between mortality and SBP using a Cox regression model adjusted for the other risk factors.

The data contained a lot of covariates (351 variables). For the imputation, we used 10 covariates that had correlations with SBP (above 0.15). These covariates were sex, age, serum albumin, Cognition (mini-mental-state examination), current hypertension, serum cholesterol, fraction erythrocytes, haemoglobin, haematocrit, and treated by specialist. For the nonresponse model, in addition to SBP, we included extra covariates that were potentially related to the nonresponse such as year of interview, activities of daily living, and so on. It should be noted that few covariates were incomplete, and accounting for them would introduce additional complexity into the imputation algorithm because of the multivariate nature. As a temporary fix, we separately imputed the missing values in the covariates and considered them as if they were real.

\subsection{Result}
\begin{figure}[!h]
\centering
{\footnotesize
\includegraphics[scale=.75, trim = 0mm 0mm 0mm 0mm]{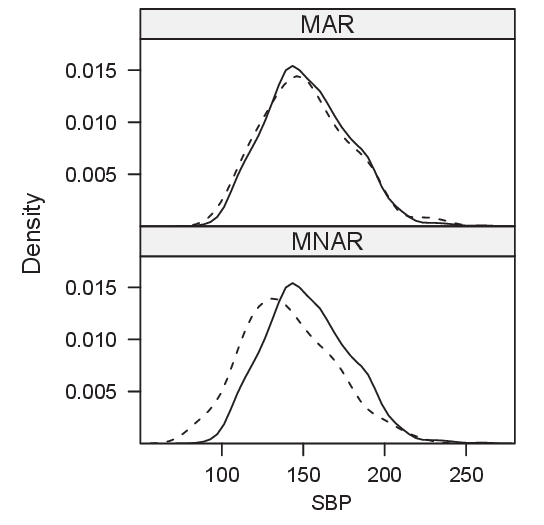}
\caption{Density plot of the observed (solid) and imputed (dashed) systolic blood pressure under MAR (MI method) and NMAR (the RI method)}
\label{leiden}
}
\end{figure}

Figure~\ref{leiden} compares the distribution of SBP for the observed and imputed data where the MI and RI methods are used. We observe that the imputed values under MAR have almost the same distribution as the observed values do (top). This is because the imputation method under MAR assumes an identical distribution for the missing and observed parts of the SBP given the other variables. On the other hand, the RI method allows for systematic differences between the missing and observed parts beyond MAR. As can be seen from the bottom of the graph, the density function of the SBP for the imputed values is quite different from that of the observed values. Interestingly, this result agrees with Table IV of \shortciteA{buuren99} where they presented a numerical example of a rather extreme MNAR mechanism with more missing values for lower SBP.

Table~\ref{syst} shows the mean of the SBP together with its standard error for the whole sample (the observed and imputed values of the SBP). The mean of the imputed values for the SBP (with its standard error) is also presented separately. We compare three methods: CC, MI, and RI. As we expected, the total mean of the SBP is lower in the RI than the other methods. Moreover, comparing the mean of the imputed values indicates that 10 mmHg difference (149.5 - 139.1 $\approx$ 10 mmHg) is caused by the nonignorable aspect. It confirms what we know about the collection process. The average amount of the adjustment used by the RI method can be calculated by $\d$ = 152.9 - 139.1 = 13.8 mmHg. This adjustment is similar to the amount of the adjustment in \shortciteA{buuren99} based on a numerical example. This results suggest that the CC and MI overestimate the mean of the SBP.

\begin{table}[!h]
\centering
{\footnotesize
\caption{Mean and standard error (SE) for the systolic blood pressure using CC, MI and RI methods.}
\label{syst}
\smallskip
\begin{tabular}{lccccc}
  \hline\hline
         &\multicolumn{2}{c}{Total} && \multicolumn{2}{c}{Imputed}\\
 \cline{2-3}\cline{5-6}  Method & Mean & SE && Mean & SE\\

  \hline
  CC     & 152.893 & 0.892 && -      & -     \\
  MI     & 152.473 & 0.924 && 149.47 & 2.409 \\
  RI     & 151.075 & 1.109 && 139.06 & 2.438 \\
  \hline\hline
\end{tabular}
}
\end{table}

After imputing the missing values in the SBP, we fit a Cox regression of mortality on the SBP adjusted for age, sex, and health measurements. As we are only interested in the effect of the SBP on the survival time, the raw estimate of its coefficient, which represents the effect of a unit difference in SBP on log hazard function, is shown here. Since a unit difference is not meaningful because of very small raw estimate, we use a standard deviation unit difference (standard deviation = 25.76). The estimated coefficient of SBP (and its standard error in parentheses) by the RI method is -0.050 (0.047) which is slightly different from -0.041 (0.044) by MI and -0.039 (0.045) by CC. Although we might have expected that the RI method would give different results than the CC, the differences turned to be small for several reasons. First, the prediction of the SBP was relatively poor ($R^2 = 0.17$), so the imputations contain considerable residual noise. Also, the missingness rate was small ($12.5\%$).

In sum, it appears that the RI method behaves as intended. It is worth noting that the SBP was suspicious of being MNAR, so the RI method has a potential advantage by allowing a systematic difference between the missing part of SBP and the observed part of it.

\section{Discussion}
The current implementations of MI assume MAR, though it can be used in combination with sensitivity analysis for MNAR data. Under MNAR, MI requires the user to specify parameters that cannot be tested on the data. The RI method, in contrast, estimates these parameters from the data, and is thus automatic without requiring user intervention. Our new method is a natural and general extension of the class of imputation methods to data that are MNAR. Drawing random indicators is general, and easy to implement in existing software. 

The results of this study suggest that the RI method can be a good alternative of dealing with MNAR data, both when the nonresponse model is related exclusively to $X_1$, and in combination with a complete covariate $\Z$. Of course, when the complete covariate $\Z$ is available, it is wise to include this information to imputation model.

The price of the RI method is an increase in standard error. The method provides slightly larger standard errors than the complete case method. Also note that the variances of the observed and missing parts of the incomplete variable are assumed to be equal. This assumption is needed in order to be able to estimate a shift in the mean of the missing data. Though this assumption is a usual one in statistics, we should be cautious in using the RI method if the variances are clearly different.

The RI method assumes a parametric model for the missingness mechanism, as a logistic link for the response process. Other types of modeling of the missingness mechanism such as semi-parametric and non-parametric methods can also be used within the RI framework to draw imputations under MNAR. Unfortunately, these modeling assumptions cannot fully be verified from the data because of the missing values. Thus, further research is needed to check the sensitivity of the results on the different models for the missingness mechanism.


The RI technique can be refined in various ways. This study was restricted to univariate missing data. A logical extension of the RI method is the analysis of multivariate missing data where several variables have missing values. This multivariate extension can be done by two general strategies for imputing multivariate missing data: joint modeling and fully conditional specification \shortcite{buuren06}. Another extension would be to increase the number of draws from the nonresponse model. Here, we generated one realization of the response indicator. Using multiple realizations of the response indicator yields a more refined way to decompose the missing part of the incomplete variable. Such an extension allows for a much closer modeling of the probabilities of the nonresponse mechanism. Generalizations like these will contribute to a fruitful line of imputation methods for data that are missing not at random.

\bibliographystyle{apacitex}
\bibliography{jolani}

\section*{Appendix A}
\appendix

\subsection{Proof of Proposition 1}
The density function of $R$ given $x_1$ and $z$ is defined by
\begin{eqnarray*}
f(R|x_1,z)=\frac{e^{R(\psi_0 + \psi_1x_1 + \psi_2z)}}{1+e^{\psi_0 + \psi_1x_1 + \psi_2z}} \qquad R = 0, 1.
\end{eqnarray*}
By substituting $R = 1$ in the above equation, we have
$$f(R = 1|x_1,z)=\frac{e^{(\psi_0 + \psi_1x_1 + \psi_2z)}}{1+e^{\psi_0 + \psi_1x_1 + \psi_2z}}.$$
Thus, we can rewrite the density function $f(R|x_1,z)$ by
\begin{eqnarray}
\label{conjo}
f(R|x_1,z)=e^{(\psi_0 + \psi_1x_1 + \psi_2z)(R - 1)}f(R = 1|x_1,z) \qquad R = 0, 1.
\end{eqnarray}

The marginal distribution function $f(R|z)$ is obtained as follows:
\begin{eqnarray*}
f(R|z) &=& \int f(R, x_1|z)\partial x_1\\
               &=& \int f(R|x_1,z)f(x_1|z)\partial x_1\\
               &=& \int e^{(\psi_0 + \psi_1x_1 + \psi_2z)(R - 1)}f(R - 1|x_1,z)f(x_1|z)\partial x_1\\
               &=& f(R = 1|z)\int e^{(\psi_0 + \psi_1x_1 + \psi_2z)(R - 1)}\frac{f(R = 1|x_1,z)f(x_1|z)}{f(R = 1|z)}\partial x_1\\
               &=& f(R = 1|z)\int e^{(\psi_0 + \psi_1x_1 + \psi_2z)(R - 1)}f(x_1|z,R = 1)\partial x_1\\
               &=& f(R = 1|z)e^{(\psi_0 + \psi_2z)(R - 1)}\int e^{\psi_1(R - 1)x_1}f(x_1|z,R = 1)\partial x_1\\
               &=& f(R = 1|z)e^{(\psi_0 + \psi_2z)(R - 1)}\textbf{\textit{M}}_{x_1|z,R = 1}[\psi_1(R - 1)]\\
\end{eqnarray*}
where $\textbf{\textit{M}}$ is the moment generating function of $x_1$ given $z$ and $R = 1$. Suppose $f(x_1|z, R = 1)\sim N(\mu_z, \s)$, then we can define the marginal distribution function of $R$ by
\begin{eqnarray}
\label{jo}
f(R|z) = f(R = 1|z)e^{(\psi_0 + \psi_2z)(R - 1)}e^{\mu\psi_1(R - 1)+\frac{1}{2}\psi^2_1(R - 1)^2\s}
\end{eqnarray}
for $R = 0, 1$. Now, by using Equation~\ref{conjo}, the joint distribution function of $(R, x_1|z)$ is given by
\begin{eqnarray*}
f(R, x_1|z) &=& f(R|x_1,z)f(x_1|z)\\
            &=& e^{(\psi_0 + \psi_1x_1 + \psi_2z)(R - 1)}f(R = 1|x_1,z)f(x_1|z)\\
            &=& f(R = 1|z)e^{(\psi_0 + \psi_1x_1 + \psi_2z)(R - 1)}f(x_1|z, R = 1).
\end{eqnarray*}

Finally, the conditional distribution function $x_1$ given $R$ and $z$ is defined by dividing the joint distribution function of $(R, x_1|z)$ by the marginal distribution function of $(R|z)$ (Equation~\ref{jo}), that is,
\begin{eqnarray*}
f(x_1|z, R) &=& \frac{f(R,x_1|z)}{f(R|z)}\\
            &=& f(x_1|z,R = 1)e^{(x_1-\mu_z)\psi_1(R - 1)-\frac{1}{2}\psi_1^2(R - 1)^2\s}.
\end{eqnarray*}
We emphasize that $\psi_2$ is canceled out by dividing $f(R,x_1|z)$ by $f(R|z)$. For normal $x_1$ we have $f(x_1|z, R = 1) = \frac{1}{\sqrt{2\pi\s}}e^{-\frac{1}{2\s}(x_1 - \mu_z)^2}$. It is easy to see
\begin{eqnarray*}
f(x_1|z,R) &=& \frac{1}{\sqrt{2\pi\s}}e^{-\frac{1}{2\s}[x_1 - (\mu + \psi_1(R - 1)\s)]^2} \qquad R = 0, 1,
\end{eqnarray*}
so that
$$X_1|Z, R \sim N(\mu + \psi_1(R - 1)\s, \s) \qquad R, = 0, 1,$$
and the proof is complete. $\hspace{8.5cm}\blacktriangle$

\subsection{Equivalence between $\mu_{10}$ and $\mu_{01}$}
Assume $X_1$ is a random variable and $R$ and $\dot{R}$ are two independent Bernoulli variables. We start by the definition of $\mu_{10}$:
\begin{eqnarray*}
\mu_{10} = E(X_1|R = 1, \dot{R} = 0) &=& \int x_1 P(x_1|R = 1, \dot{R} = 0)\partial x_1\\
                        &=& \int x_1\frac{P(R = 1, \dot{R} = 0|x_1)P(x_1)}{P(R = 1, \dot{R} = 0)}\partial x_1\\
                        &=& \int x_1\frac{P(R = 1|x_1) P(\dot{R} = 0|x_1)P(x_1)}{P(R = 1)P(\dot{R} = 0)}\partial x_1
\end{eqnarray*}
Because $R$ and $\dot{R}$ have the same probability function, it is obvious that $P(R = 1) = P(\dot{R} = 1)$ and $P(R = 1|x_1) =  P(\dot{R} = 1|x_1)$. Consequently,
\begin{eqnarray*}
E(X_1|R = 1, \dot{R} = 0) &=& \int x_1\frac{P(\dot{R} = 1|x_1) P(R = 0|x_1)P(x_1)}{P(\dot{R} = 1)P(R = 0)}\partial x_1\\
                        &=& \int x_1\frac{P(\dot{R} = 1, R = 0|x_1)P(x_1)}{P(\dot{R} = 1, R = 0)}\partial x_1\\
                        &=& \int x_1 P(x_1|\dot{R} = 1, R = 0)\partial x_1\\
                        &=& E(x_1|\dot{R} = 1, R = 0) = \mu_{01}
\end{eqnarray*}
and the proof is complete. $\hspace{8.5cm}\blacktriangle$

\end{document}